\documentclass[apjl]{emulateapj}
\usepackage{tipa}
\usepackage[figuresright]{rotating}
\usepackage{subfigure}
\usepackage{epic,eepic}
\usepackage{graphicx}
\usepackage{longtable}
\usepackage{threeparttable}
\usepackage{appendix}
\usepackage{amsmath}
\shorttitle{Do long-cadence data of {\it Kepler} satellite capture
the basic properties of flares ?}

\shortauthors{Yang et al.}

\begin{document}

\title{Do long-cadence data of the {\it Kepler} satellite capture basic properties of flares ? }

\author{Huiqin Yang\altaffilmark{1,2}, Jifeng Liu\altaffilmark{1,2},  Erlin Qiao\altaffilmark{1,2}, Haotong Zhang\altaffilmark{1}, Qing Gao\altaffilmark{1}, Kaiming Cui\altaffilmark{1,2},Henggeng Han\altaffilmark{1,2}}

\altaffiltext{1}{Key Laboratory of Optical Astronomy, National
Astronomical Observatories, Chinese Academy of Sciences, Beijing
100101, China; yhq@nao.cas.cn}

\altaffiltext{2}{University of Chinese Academy of Sciences, Beijing
100049, China}

\begin{abstract}
Flare research is becoming a burgeoning realm in the study of
stellar activity due to the launch of {\it Kepler} in 2009. {\it
Kepler} provides data with two time resolutions, i.e., the
long-cadence (LC) data with a time resolution of 30 minutes and the
short-cadence(SC) data with a time resolution of 1 minute, both of
which can be used to study stellar flares. In this paper, we search
flares in light curves with both LC data and SC data, and compare
them in aspects of the true-flare rate, the flare energy, the flare
amplitude, and the flare duration. It is found that LC data
systematically underestimated the energies of flares by 25\%, and
underestimated the amplitudes of flares by 60\% compared with SC
flares. The duration are systematically overestimated by 50\%
compared with SC flares. However, the above percentages are poorly
constrained and there is a lot of scatter. About 60\% SC flares have
not been detected by LC data. We investigate the limitation of LC
data, and suggest that although LC data cannot reflect the detailed
profiles of flares, they also can capture the basic properties of
stellar flares.
\end{abstract}

\keywords{Stars:flare --- methods: analytical --- methods:
statistical}

\section{INTRODUCTION}
The {\it Kepler} satellite \citep {Borucki2010} has yielded the
light curves of more than 200,000 stars with unprecedented precision
up to 17 quarters of continuous observations. Its precision for
bright targets (V = 9-10) approaches 10 ppm, and 100 ppm for fainter
targets(V = 13-14). It thus has ushered in a new era of stellar
photometric investigation. Using the data of {\it Kepler} mission to
study the property of the flare can reveal many very important
aspects of stellar physics, such as stellar activity, the mechanism
of superflares, stellar structure, the interaction of binary and so
on \citep [e.g.,][]
{Shibayama2013,Balona2013,Hawley2014,Gaul2014,Wich2014,Balona2015,Lurie2015,Davenport2016,Gao2016,Chang2017,Door2017,Yang2017}.

{\it Kepler} supplies two kinds of data with different time
resolution, i.e., the long-cadence (LC) data with a time resolution
of 30 minutes and the short-cadence(SC) data with a time resolution
of 1 minute sampling interval. As known on our sun, most flares are
microflares and nanoflares, which last for several minutes
\citep[e.g.,][]{Benz2010}. LC flares often have longer duration and
much more energies. Most of them thus are superflares with a low
resolution.

Obviously, SC data are more suitable for studying flares. However,
there are only about 5000 sources with SC data, and usually with
rather short time coverage (about two months on average). Therefore,
some work began to use LC data to study flares \citep [e.g.,][]
{Walkowicz2011,Shibayama2013,Davenport2016,Gao2016,Yang2017},
although deviations and errors of the data, such as the estimations
for the energies, the durations, the amplitudes, and the true-flare
rate, are not known accurately.

The differences between LC flares and SC flares may be
underestimated by previous works. For example, \citet{Shibayama2013}
compared two flares both in LC and SC data, and obtained a good
consistency of them. Then they concluded that the flare energies
estimated from LC data and SC data were similar. However, only two
examples are not enough to indicate the correspondence of LC flares
and SC flares. Actually, flares of LC and SC data may be much
different. Issues such as the misaligned peak, the profile and the
flare number can result in large deviations, which will be discussed
in this work.

\citet{Hawley2014} suggested that, in SC data, the flattening
tendency of the flare frequency distribution (FFD) is not due to
detection limit but a real feature. On the other hand,
\citet{Yang2017} compared the FFD between LC and SC data, and found
that the flattening tendency of LC data is due to detection limit.

\citet{Balona2012,Balona2013} reported flares detected in A-type
stars. This discovery has raised a big challenge to the dynamo
theory, because these stars are traditionally considered without
significant convective envelopes. However,  only LC are observed for
most of those stars, and a large fraction of flares in those LC data
are unresolved. \citet{Pedersen2017} presented detailed analyses of
those stars and found more than half of them are subject to
contaminations. It raises the question whether the analysis of
flares by LC data is reliable, given its low time resolution as
compared to SC data, which we believe can capture the detailed
properties of flares. To answer the above question, a strict
comparison of flares both in LC and SC data is essential.

Since there are many interesting results available on flares that
either support or challenge the traditional theories, it is
necessary to conduct an elaborate study on {\it Kepler} data. In
this work, we explore the relation of the flares between LC and SC
data, and carry out statistical comparison on their energies,
durations and amplitudes. We try to give quantized indices to
describe deviations and errors of LC flares, and search for
conditions where flares can be studied with LC data. In Section 2,
we introduce the data used and describe the method of analysis. In
Section 3, the main results and the statistical analyses are
presented and discussed. We present our conclusions in Section 4.

\section{Data and Method}

\begin{figure*}[!htb]
\center
\subfigure{\includegraphics[width=0.9\textwidth]{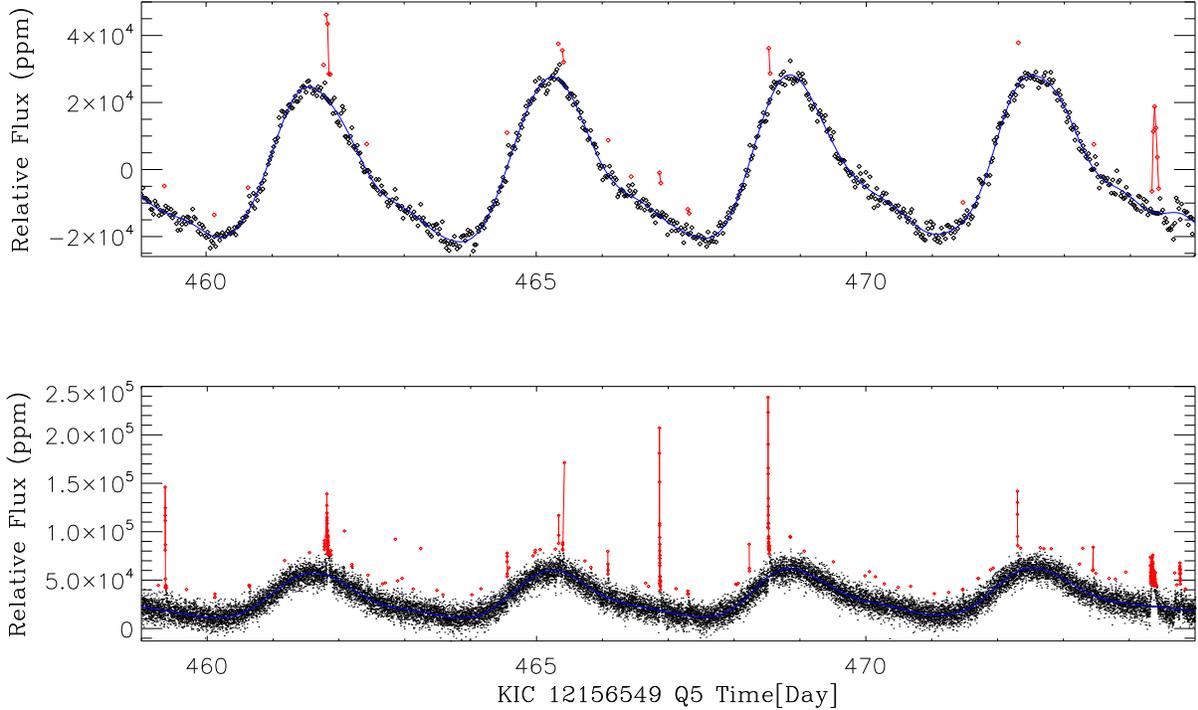}}
\hspace{1 mm} \caption{Examples of flare searching in LC and SC data
within the same time range. The top panel and bottom panel
corresponds the LC and SC observations respectively. The horizontal
axis is the {\it Kepler} Barycentric Julian Day (BKJD)
(BJD$-$2454833) and the vertical axis is the relative flux (ppm).
The black points are the observed stellar flux, the blue line are
baselines, and the red lines mark the flare candidates.}\label{fig1}
\end{figure*}

\subsection{LC and SC Data}
In the {\it Kepler} data, a set of coadded and stored pixels
obtained at a specific time is referred to as a cadence, and the
total amount of time over which the data in a cadence is coadded is
the cadence period. The two cadence periods in use are LC and SC.
Each cadence consists of a series of frames that each includes a
6.02 s exposure time and a 0.52 s readout time. For LC data and Full
Frame Images (FFI), 270 frames are coadded, leading to a total of
1765.5 s = 0.4904 h. For SC data, 9 frames are coadded, leading a
total of 58.85 s \citep{Van Cleve2009}.

Our work is based on the entire {\it Kepler} mission data set
(Q1-Q17; 48 months; Data Release 25). {\it Kepler} provides
uncorrected simple aperture photometry (SAP) and pre-search data
conditioning (PDC) in which instrumental effects are removed.
However, PDC data could also remove some outliers that may be flare
peaks. Therefore, SAP data are utilized in our research, as is done
in \citet{Balona2015} and \cite{Davenport2016}.

In total, there are 5140 targets in {\it Kepler} data with SC
observations. All of the time segments of SC data are also covered
by the LC data, i.e, all of SC data have their corresponding LC data
with the same time coverage. The flare search is carried out in
overlaps of LC and SC data.

\subsection{Flare detection and Energy Estimation}
The methods of flare detection and energy estimation are the same as
those used in \citet{Yang2017} and we refer the interested readers
to this reference for a detailed description. Here, we only present
the main steps to detect flares: (1) To fit the baseline with an
appropriate median filter. An iterative $\sigma$-clipping approach
is applied to remove possible flares before fitting the continuum.
(2) All outliers of LC data above $3\sigma$ against the baseline are
marked as flare candidates, while in SC data, outliers with no less
than three consecutive points are defined as flare candidates.

Figure~\ref{fig1} illustrates the flare detection, and gives a
comparison of flare candidates between LC and SC data. It could be
inferred that amplitudes of the flares with different time
resolution are not correlated well. Many small LC flares show high
amplitudes in SC data. Most of the isolated outliers with only one
point of LC data are artifacts, while a few of them correspond true
flares in SC data.

For a detailed description of our approach for the energy
estimation, the readers are referred to \citet{Shibayama2013} and
\citet{Yang2017}. Its main principles are as follows: (1) The white
light flare can be described by a black-body radiation model with
the effective temperature of 9000 K. (2) The flare area can be
calculated by flare amplitude, stellar radius, stellar effective
temperature and {\it Kepler} response function.
\begin{figure}[!htb]
\center
\subfigure{\includegraphics[width=0.45\textwidth]{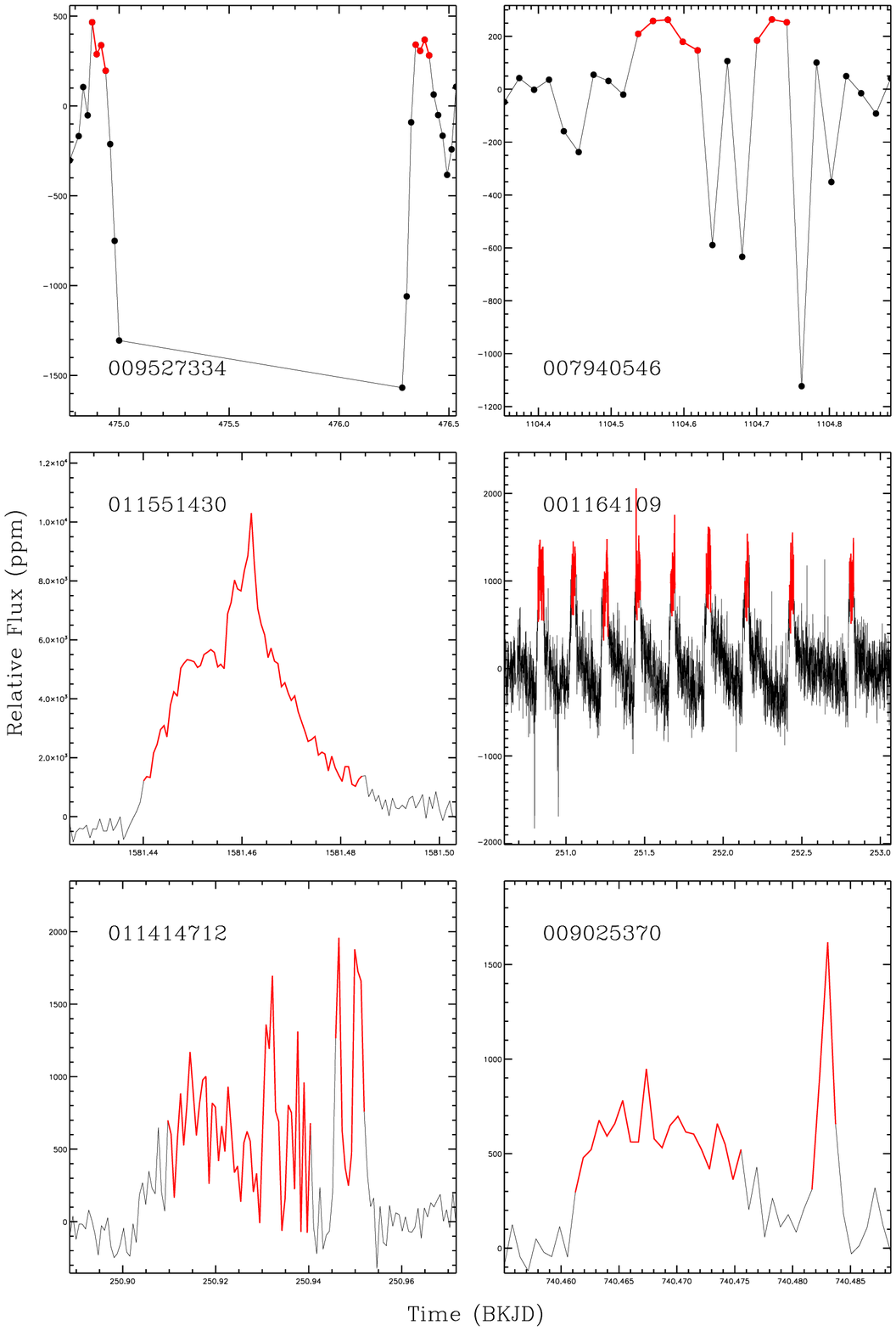}}
\hspace{2 mm}
\caption{ Examples of various artificial flares in LC (top) and SC
(middle and bottom) data. The black line are the detrended flux. The
red lines are the false positive flares. The top left panel shows
the artifact caused by discontinuity. The top right panel shows the
abnormally low points, which could cause their nearby points to be
higher than $3\sigma$. The middle left panel shows the symmetrical
profile of a flare candidate, which is removed in our sample. The
middle right panel shows the periodically impulsive signals in Q2.
The bottom panels show the false positive signals that appear in the
{\it Kepler} records for many stars at the same time segments.}
\label{fig2}
\end{figure}

\subsection{What is a True Flare?}

For most flares, the time resolution of LC data is low, of which
even an isolated outlier may be a true flare. The time resolution of
SC data is enough. However, flares in SC data could also be
unreliable. For example, Figure~\ref{fig2} shows various
false-positive signals in SC and LC data. A symmetric profile
(middle left panel), a periodically impulsive artifact (middle right
panel), or even missing data could cause false-positive signals,
which imply instrumental errors or other artifacts.

As our purpose is to compare flares of LC and SC data, here, we
stipulate that a true flare need to meet three basic criteria:

(1) In LC and SC data, the corresponding flares should occur at the
same time within a window of half an hour.

(2) The SC flare should have at least three consecutive points,
which are 3$\sigma$ above the baseline and is fully resolved.

(3) The profile of flares in SC data should have an impulsive rise
and an exponential decay, i.e., the decay duration should be longer
than the rise duration. For instance, the middle left panel of
Figure~\ref{fig2} shows the example of a symmetric flare candidate.
It should be noted that a lot of solar flares may have shapes that
differ from the assumed profile.

\begin{table}
\renewcommand{\arraystretch}{1.5}
\begin{center}
\caption[]{ Time segments containing discontinuities or abnormal low
points. }
 \label{tablerror}\small
\tabcolsep=4.pt
\begin{threeparttable}
\begin{tabular}{cccccccccc}
\tableline \tableline Quarter & Begin Time
& End Time & Begin Time & End Time\\

& (Day)  & (Day) & (Day) & (Day)\\
\hline

Q2 &250.80& 251.00 &--&--\\
Q3 &333.30& 333.50 &342.50& 342.90\\
Q7 &633.70& 634.30 &644.20& 644.50\\
Q7 &659.10& 659.15 &--&--\\
Q8 &737.30& 737.60 &740.40&740.45\\
Q9 &850.20& 850.40 &929.50&929.80\\
Q10&946.00& 946.30 &--&--\\
Q11&1003.20&1004.60&1008.20&1008.50\\
Q11&1021.30&1021.50&--&--\\
Q14&1278.50&1278.70&--&--\\
 \tableline

\end{tabular}
    \begin{tablenotes}
       \item[] Notes: Instrumental errors are numerous and pervasive
       in the above time segments.
       All the flare candidates in them are removed from our sample.
      \end{tablenotes}
\end{threeparttable}
\end{center}
\end{table}
We emphasize that the purpose of the above three criteria is mainly
to filter out artifacts rather than to identify true flares. In
fact, there are various artifacts in the {\it Kepler} data that meet
all the three criteria above. For example, the top panels of
Figure~\ref{fig2} show the artifacts caused by abnormal or missing
points. Some work mentioned those issues
\citep[e.g.,][]{Pedersen2017}, but none has a comprehensive study of
these artifacts. We checked some target-pixel-files of the
abnormally low points, and found the counts of some pixels decreased
rapidly at some time, while other pixels increase dramatically
around the same time. \citet{Coughlin2016} reported ``column
anomaly" in {\it Kepler} data, which is due to decreasing charge
transfer efficiency over time. It seems that those high points and
low points are caused by the similar reasons.

Moreover, they are numerous and pervasive, and some of them occur in
the same time segments. We believe that they are artifacts, and
remove most of the flare candidates with discontinuities and all the
flare candidates with the abnormally low points nearby.
Table~\ref{tablerror} shows the time segments which often have
discontinuities or abnormally low points sometimes with similar
shapes as shown in the top panels of Figure~\ref{fig2}. More than
1,300 flare candidates are removed due to this reason. The middle
right panel of Figure~\ref{fig2} illustrates periodically impulsive
signals, which happen in a well-known segments of unreliability: Q2
\citep[e.g.,][]{Coughlin2016}.

Physical parameters such as temperatures are given by the Kepler
Input Catalog \citep[KIC;][]{Brown2011,Huber2014}. Eight stars in
our sample exceed 10,000K. They are white dwarfs, polar, hot
subdwarfs and planetary nebula. The large deviations of KIC on early
stars are frequently reported
\citep[e.g.,][]{Dressing2013,Batalha2013,Huber2014}, and are
supported by work on compact stars. For example, KIC 11822535 is a
DA type white dwarf with $T_{\ast}$ ${\sim}$35\,000~K
\citep{Gian2011,Barstow2014}, but its temperature in KIC is about
10,000~K. All of the stars whose temperature exceed 10,000~K are
excluded.

It should be noted that the adopted flare detection procedure has
several stages with heuristic choices that are based on our
experience. All the final flares are identified by visual
inspection. Although the resulting flare catalog is not completely
reproducible, the criteria are clear, and the results are objective.
\begin{table}
\renewcommand{\arraystretch}{1.5}
\begin{center}
\caption[]{ The parameters of LC flares }
 \label{tablelc}\small
\tabcolsep=5.pt
\begin{threeparttable}
\begin{tabular}{cccccccccc}
\tableline \tableline KIC & Ind.
& BTime & ETime & Amp. & Energy \\

&& (Day)  & (Day) & (ppm) & (erg)\\
\hline

1025986& 0 & 294.2454 & 294.2454   &  288.8&32.63 \\
1570924&    1& 1193.4238 &1193.4851   & 5384.0&33.95 \\
1570924 &  2 &1193.8733& 1193.8938   & 7930.9&33.60  \\
1570924   & 3 &1194.6294 &1194.6703   & 6659.5&33.87 \\
1570924  & 4 &1206.4406 &1206.4406   & 3234.9&33.26  \\
2300039 & 5 & 909.6052 & 909.6052   & 4514.5&31.55 \\

 \tableline

\end{tabular}
    \begin{tablenotes}
       \item[] Notes: Index connects the flare between LC and SC
       data. A LC flare could correspond multiple SC flares. Energy are in logarithm.

       (This table is available in its entirety in machine-readable form.)
      \end{tablenotes}
\end{threeparttable}
\end{center}
\end{table}

\begin{table}
\renewcommand{\arraystretch}{1.5}

\caption[]{ The parameters of SC flares }
 \label{tablesc}\small
\tabcolsep=5.pt
\begin{threeparttable}
\begin{tabular}{lllllllll}
\tableline \tableline KIC & Ind.
& BTime & ETime & Amp. & Energy \\

&& (Day)  & (Day) & (ppm) & (erg)\\
\hline

1025986 &0&294.2396&294.2417&1122.1&32.44\\
1570924 &1&1193.4262& 1193.4623&10204.0&33.92\\
1570924 &1&1193.4718& 1193.4732&4697.4&32.29\\
1570924 &2&1193.8703& 1193.8866&30803.9&33.93\\
1570924 &3&1194.6311& 1194.6774&10640.2&34.03\\
1570924 &4&1206.4362& 1206.4437&12974.9&33.27\\
2300039 &5& 909.6096& 909.6116&30496.6&31.57\\

 \tableline

\end{tabular}
\begin{tablenotes}
\item[] Notes: Same as Table~\ref{tablelc}, but for SC flares.
       (This table is available in its entirety in machine-readable form.)
\end{tablenotes}
\end{threeparttable}

\end{table}

\subsection{Contamination Check}
As the typical photometric aperture of {\it Kepler} has a radius of
4--7 pixels \citep{Bryson2010}, it is thus quite common for the
observations of a given target star to be contaminated by nearby
objects, given that some sources are very close to each other on
{\it Kepler}'s CCDs. About 10\% flare candidates are probably untrue
due to various reasons \citep{Shibayama2013, Gao2016}. It is thus
necessary to check contamination. The contamination check is similar
with \citet{Yang2017}, which can be summarized into three aspects as
follows:

(1) 45 flaring stars with field stars located within 12\arcsec\ are
excluded \citep{Maehara2012}.

(2) The {\it Kepler} eclipsing binary catalog
(KEB)\footnote{http://keplerebs.villanova.edu/} (released on 2017
September 20) includes more than 2800 eclipsing binaries. 56 of them
are removed from our sample.

(3) The checks of centroid offset are done \citep{Yang2017}. About
one hundred flares are removed in this step.
\section{Results}

Our aims are to compare properties of flares between LC and SC data,
to explore the conditions of trusting LC flares, and to give
corrections and errors of LC flares. Thus, our work is based on LC
flares, and SC flares are deemed as standard. In total, 30,485 flare
candidates above $3\sigma$ are found in LC data, in which 940 are
true flares. All the true flares meet the above criteria, and have
been checked by eye. For LC flares, Table~\ref{tablelc} gives their
start time, end time, peaks and energies. For SC flares,
Table~\ref{tablesc} gives their start time, end time, peaks and
energies. The appendix provides images for each flare, which plot LC
and SC data together.

On the other hand, it is interesting to investigate how many SC
flares have not been detected in LC data. This is important in terms
of determining how well the frequency of flares can be represented
by LC data. \citet{Balona2015} reported 3,140 flares in SC data from
Q1 to Q12 by visual inspection, and his results are without
contamination check. In this work, 3,878 SC flares are identified
through the entire data set after removing about 100 contaminated
stars. 1,425 SC flares have their counterparts, and 2,453 (63\%) SC
flares have been lost by LC data. However, it should be \text{noted}
that the loss rate is an average. Since the FFDs vary intensely in
different stars \citep{Davenport2016}, the loss rate may be much
different for individual star. The analysis for each star is beyond
the scope of this work.

Table~\ref{tablefn} lists the comparison of detecting flares
in LC and SC data. It illustrates the process of our detection.
Flare candidates are independently detected in each mode and are
without any additional check. True flares are results after
pollution check and profile check. It should be noted that the flare
correspondence between different modes is complex. Given the low
resolution of LC data, a LC flare may not resolve multiple flares in
corresponding SC data, which occurred at a very close time. Strictly
speaking, each SC flare is a part of the LC flare, whereas all of
them correspond to the same LC flare according to the definition of
Section 2.3. One can see many examples of this case in the appendix
images. We hereby refer counterpart as a extended meaning, which
reflects that a LC flare may have multiple SC counterparts. Hence
940 LC flares correspond to 1425 SC flares.
%
%
%
%
\begin{table}
\renewcommand{\arraystretch}{1.5}
\begin{center}
\caption[]{ Comparison of detecting flares in different mode}
 \label{tablefn}\small
\tabcolsep=14.pt
\begin{threeparttable}
\begin{tabular}{cccccccccc}
\tableline \tableline
Mode & FC\tnote{a} & CC\tnote{b}& TF\tnote{c} & TFC\tnote{d}\\
\hline

LC &30485&3266&940&940 \\
SC &7397&4762&3878&1425\\

 \tableline

\end{tabular}
    \begin{tablenotes}
       \item[] Notes: Flare candidate represents candidates without any additional
       check. True flare represents results after pollution check and profile check.
       If a flare has a counterpart, it means that its counterpart should occur at the
      same time within a window of half an hour in its corresponding
       mode. Note that one LC flare may have multiple SC counterparts
       because of its low resolution. One can see an example in the
       right panel of the appendix.
     \item[a] Flare candidate.
     \item[b] Candidate with counterpart.
     \item[c] True flare.
     \item[d] True flare with counterpart.
      \end{tablenotes}
\end{threeparttable}
\end{center}
\end{table}

\begin{table}
\renewcommand{\arraystretch}{1.5}
\begin{center}
\caption[]{ True-flare rate in LC data }
 \label{tableac}\small
\tabcolsep=5.pt
\begin{threeparttable}
\begin{tabular}{cccccccccc}
\tableline \tableline
Number of Points & True Flares & Candidates  & True-flare Rate \\
\hline

One point &534& 29808&1.79\% \\
Two points&141& 332&42.47\%\\
Three points &87 &124 &70.16\%\\
Four points &58& 82&70.73\% \\
Five points &35& 45&77.78\% \\
Five points above &85& 94&90.42\%\\
Total points &940&30485&3.08\%\\

 \tableline

\end{tabular}
    \begin{tablenotes}
       \item[] Notes: The detection accuracy for flares with different
     number of points in LC data.
      \end{tablenotes}
\end{threeparttable}
\end{center}
\end{table}

\subsection{True-Flare Rate}
Statistic research on flares is an important realm of stellar
activity \citep{Yang2017}. It thus is interesting to study the
true-flare rate of LC flares, which could give suggestion in the
future flare research. In the 30,485 flare candidates of LC data,
after removing all kinds of pollution, 940 LC flares with 1425 SC
counterparts are left.

Table~\ref{tableac} shows the true flares, the total flare
candidates, and the true-flare rate, which are grouped by the number
of points in a LC flare. More than 95\% flare candidates are
one-point outliers, but most of them are artifacts, which are mainly
due to instrumental errors and cosmic rays. The previous work on
flare detection often require at least two consecutive points in LC
data \citep[e.g.,][]{Walkowicz2011, Osten2012, Davenport2016,
Gao2016, Pedersen2017, Door2017, Yang2017}. The result of
Table~\ref{tableac} suggests that the criteria are reasonable, but
on the other hand, they may filter out more than half of true
flares.

When flare candidates have multiple points, the detection accuracy
rises significantly. Their counterparts in SC data have classical
flare profile. However, a considerable part of those candidates have
symmetric profiles, abnormal points or discontinuities, and are not
deemed as true flares. Some of them, such as shown in the middle
left panel of Figure~\ref{fig2}, may not be artifacts, but have
other mechanisms of flux enhancement, .

\begin{figure}[!htb]
\center
\subfigure{\includegraphics[width=0.45\textwidth]{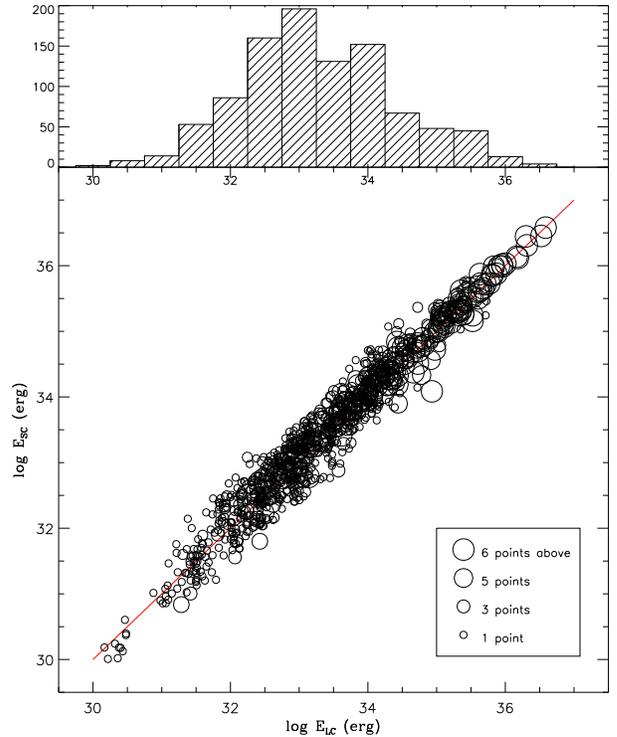}}
\hspace{2 mm}
\caption{LC energy vs. SC energy. The top panel shows the energy
distribution. The circle size indicates the number of points in LC
flares. The red line is the diagonal.} \label{figegy}
\end{figure}
\begin{figure}[!htb]
\center
\subfigure{\includegraphics[width=0.45\textwidth]{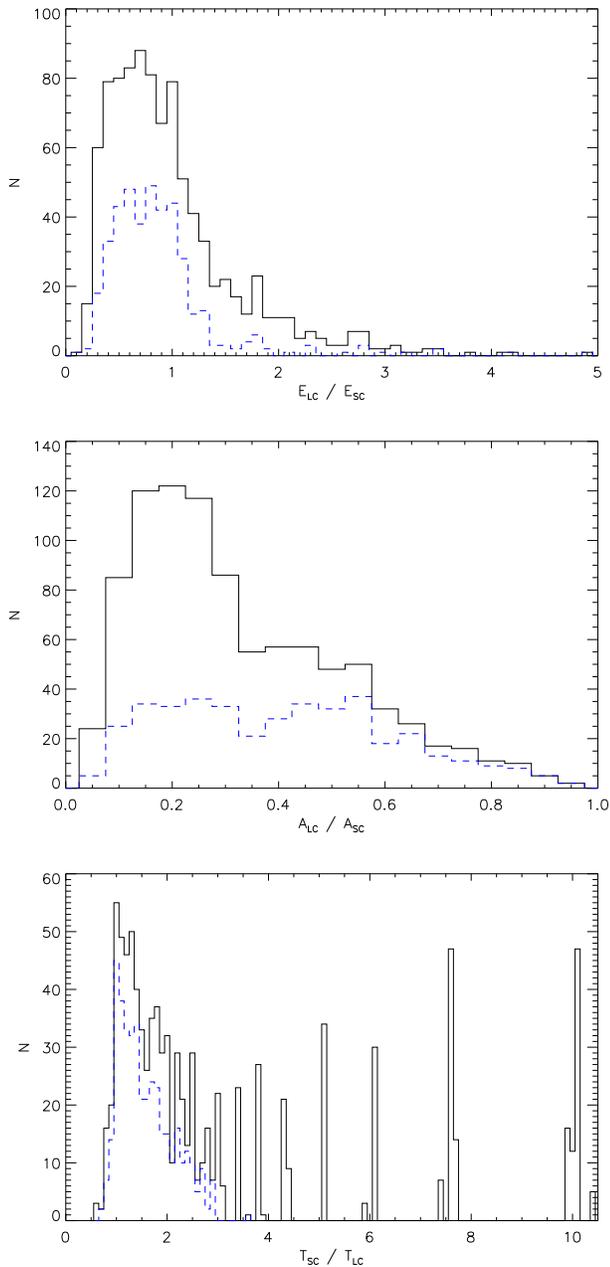}}
\hspace{2 mm}
\caption{The distribution of the ratios of the energies (top), the
amplitudes (middle), and the durations (bottom). The solid line
includes all flares, while the dashed line represents flares with
multiple points in LC data. } \label{fighist}
\end{figure}

\begin{figure}[!htb]
\center
\subfigure{\includegraphics[width=0.45\textwidth]{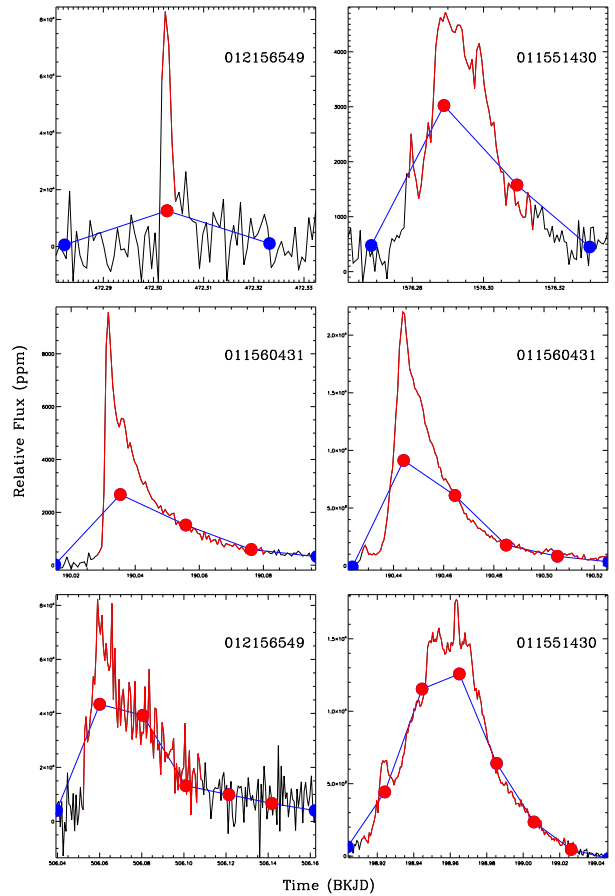}}
\hspace{2 mm}
\caption{Examples of flares, whose peaks in LC and SC data are
aligned. The black lines are the detrended flux in SC data, and the
red lines are the flares in SC data. The blue lines are the
detrended flux in LC data, and the red points are the flares in LC
data. Each panel shows the flares with different number of points in
LC data.} \label{figalign}
\end{figure}
\begin{figure}[!htb]
\center
\subfigure{\includegraphics[width=0.45\textwidth]{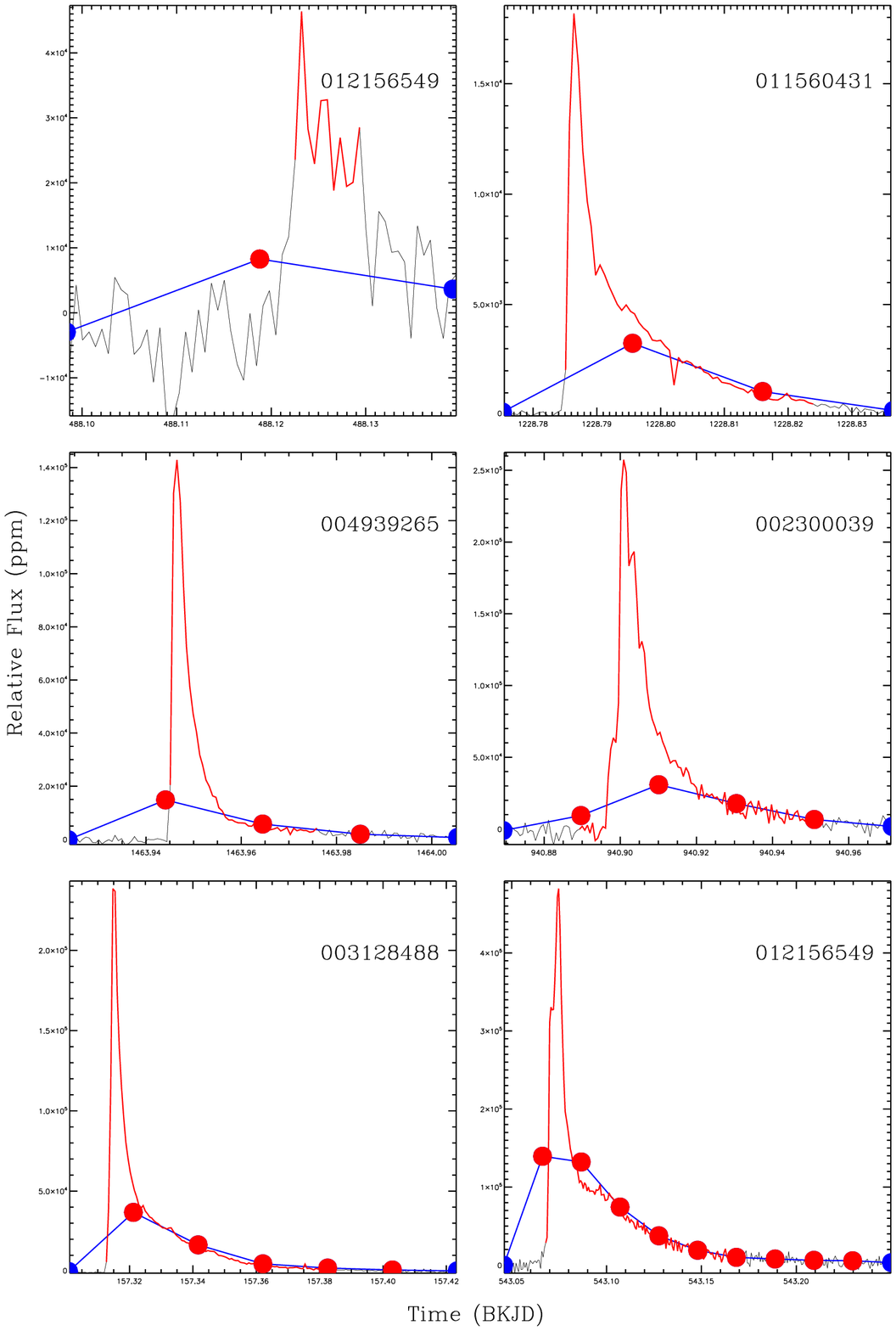}}
\hspace{2 mm}
\caption{Same as Figure~\ref{figalign} but for the flares whose
peaks are misaligned.} \label{figmisalg}
\end{figure}
\subsection{The Ratios of LC and SC Flares}
Figure~\ref{figegy} gives an overview of the energy distribution and
shows the energy relation between LC flares and SC flares. As shown
in the bottom panel, the energies are overestimated by the LC data
when the energy of the flare is low. The high energy band (above
$10^{36}$ erg) have a low dispersion. Presumably this is because
energies are proportional to durations, and the dispersions of
durations decline as the durations increase.

The profile of a flare will apparently affect the estimation of
flare parameters, while LC data are hard to capture the profiles.
Three parameters, the energy, the amplitude and the duration are
often used to quantify flare properties in the previous work
\citep{Walkowicz2011,Maehara2012,Balona2013,Shibayama2013,Cand2014,Ramsay2014,Kitze2014,Lurie2015,Gao2016,Davenport2016,Gizis2017,Makarov2017,Yang2017}.
 However, LC data are used to study flare in those work. It is thus
 necessary to study how using LC instead of SC will affect the above
 three parameters.

 Figure~\ref{fighist} shows the distribution of ratios of
energies, amplitudes, and durations between LC and SC flares. The
solid lines are the distribution of all flares, while the blue lines
represent flares with multiple points in LC data. As shown in the
top panel of Figure~\ref{fighist}, most ratios are below 1. Their
peak is near 0.8, and the concentration of the distribution is
obvious, which suggests LC data underestimate flare energies.

The middle panel of Figure~\ref{fighist} implies that one-point
flares in LC data have much larger deviations from true amplitudes.
Flares with multiple points underestimate amplitudes by 40\% to
80\%, and their distribution is wide.

The bottom panel of Figure~\ref{fighist} shows that durations of
small flares with one point are overestimated. Their real durations
are much smaller than half an hour. Flares with multiple points also
have overestimated durations. Their distribution is quite wide,
because of huge difference of the time resolution between the two
data sets. Moreover, the durations of LC flares largely depend on
baselines, which could have errors in the determination of the end
time of flares.

Therefore, from above panels, we suggest that energies and
amplitudes have been underestimated by LC flares, while LC durations
have been overestimated. For the flares with multiple points, the
mean values of these ratios demonstrate the systematical deviations
of LC flares compared to SC flares. They are 75\%, 40\% and 147\%
for energies, amplitudes and durations respectively, and their
standard deviations are 61.8\%, 21.1\% and 53.5\% respectively,
which can represent the uncertainties. The above results imply that
basic properties of flares can be captured by LC data in spite of
the existence of deviations and errors.

\subsection{The Limitation of LC data}
\begin{figure}[!htb]
\center
\subfigure{\includegraphics[width=0.45\textwidth]{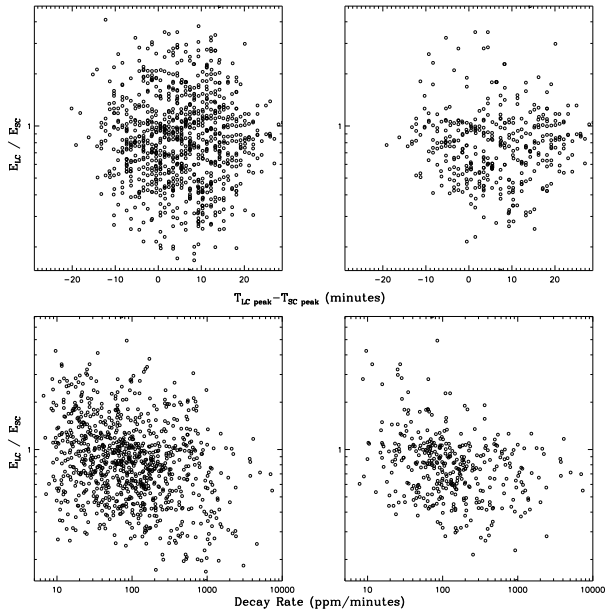}}
\hspace{2 mm}
\caption{Relation between energy ratios and parameters of flares. In
the top panels, the horizontal axis is the difference of timings
between LC and SC flares, in which the left panel shows all flares,
and the right panel shows flares with multiple points in LC data. In
the bottom panels, the horizontal axis is the decay rate.}
\label{figtcor}
\end{figure}

The deviations of basic parameters are determined by profiles of
flares. The distribution shifts of Figure~\ref{fighist} suggest that
there are differences of profiles between LC and SC data. If LC data
can capture the differences, we are able to make corrections for
each flare by LC parameters, which could provide references for the
future work.

The main energy of a flare is from its peak. However, the time
resolution of LC data is low. It is a natural thought that the
sampling timing may make great impact on the flare amplitude and the
energy estimation. Figure~\ref{figalign} illustrates some well
aligned (the difference of the sampling timings of the peaks are
within 10 minutes) LC and SC flares, while Figure~\ref{figmisalg}
shows some misaligned (the difference of the sampling timings of the
peaks are over 10 minutes) flares. Most of the peaks of LC flares
are much lower than those of SC flares, which cause 0.2 deviation
from 1. However, if the peak is low, and the profile of a flare is
flat, the LC energy also can be larger than the SC energy. The top
panels of Figure~\ref{figtcor} show the relation between energy
ratios and difference of peak timings, which imply the energy ratio
do not depend on the sampling timing, or the difference of the
sampling timing is too small to affect the ratio. Thus the influence
of the misaligned peak is limited.

Another factor that affects the flare amplitude of LC data is the
decay rate of the peak. Because the flux of each point is a mean in
a cadence period, which consists of a mount of frames. The slower
the decay rate is, the closer the flux is to the actual value. This
conclusion can be inferred from Figure~\ref{figalign} and
Figure~\ref{figmisalg}. A flare with a long duration of the peak has
a much higher amplitude than that with a short duration of the peak.
Besides, if the profile is sharp, the well-aligned peak of LC data
could also be a tenth of that of SC data.

Some previous work describe the decay rate with an exponential
function \citep[e.g.,][]{Pugh2016, Door2017}. Here, we introduce a
parameter of the LC flare: the duration from the peak to the half of
the peak (DPH), and define the decay rate as the ratio of the peak
to DHP. The DPHs are calculated by interpolation. The peak and DPH
have high signal to noise ratios and are not affected by the end
time of flares. Flares with multiple peaks are excluded in the
comparison.

The bottom panels of Figure~\ref{figtcor} show the relation between
the energy ratios and the decay rates. In the left panel, there is a
weak relation, and the dispersion is large. However, it can be
inferred that the energy estimation of LC data decreases as the peak
becomes steep.

The right panel of Figure~\ref{figtcor} shows the same as the left
panel but for the flares with multiple points. It indicates a
similar relation when the decay rates are small, and the dispersion
is large as well. However, the ratios are likely to be unchanged
when the decay rates are large enough, and the dispersion becomes
relative small. This result can also be inferred from
Figure~\ref{figegy}, which shows smaller dispersion in the high
energy band.

\begin{figure}[!htb]
\center
\subfigure{\includegraphics[width=0.45\textwidth]{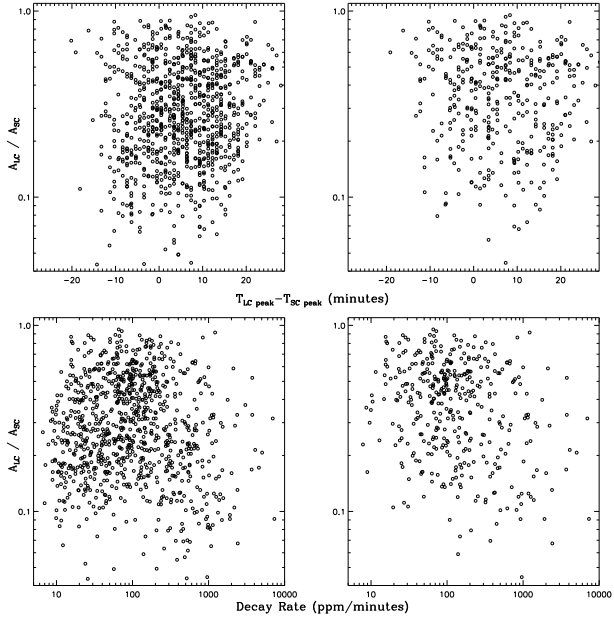}}
\hspace{2 mm}
\caption{Relation between the amplitude ratios and the flare
parameters. In the top panels, the horizontal axis is the difference
of timings between LC and SC flares, in which the left panel shows
all flares, and the right panel shows flares with at least two
points in LC data. In the bottom panels, the horizontal axis is the
decay rate.} \label{figpeakcor}
\end{figure}
 The top panels of Figure~\ref{figpeakcor} show relation between the amplitude ratios and
the difference of peak timings. Similar to Figure~\ref{figtcor}, it
implies that the influence of the misaligned peak is limited. The
bottom panels of Figure~\ref{figpeakcor} show relation between the
amplitude ratios and the decay rates. The left panel includes all
flares and shows no correlation, because one-point flares
concentrate near the low band of the decay rate and have large
errors.

However, as shown in the bottom right panel, there is a generally
weak relation with a large dispersion. The ratios decline along with
the increase of the decay rates, which demonstrates that the steeper
profiles would result in larger deviations of amplitudes.

Figure~\ref{figpeakcor} and Figure~\ref{figtcor} investigate the
limitation of LC data. The generally weak relations with large
dispersions are presented. Obviously, LC data cannot grasp the
detailed profiles of flares. It is thus hard to make correction
based on LC data, whereas the limitation of LC data may reflect the
tendency of the deviations to a certain extent.
\section{Conclusion}

The study of flare using LC data has been applied to many aspects of
stellar physics. However, the basic properties derived from the LC
data such as the true-flare rate, the energy, the amplitude, and the
duration, systematically deviate from the results derived from the
SC data.

We search flares both in LC and SC data. The search procedure and
criteria are strict, and the results are checked manually. By
comparing the LC data with the SC data, we have the conclusions as
follows:

(1) Most flare candidates with only one point in LC data are
artifacts. Few of them are true flares, while they have a large
deviation from the properties derived from SC data. Most candidates
with multiple points in LC data are true flares. However, some
candidates with symmetric profiles may not be true flares, but have
other mechanisms of flux enhancement.

On the other hand, about 60\% SC flares are lost by LC data on
average, although the loss rate may be much different for each star.

(2) LC data have an obvious underestimation on the flare energy.
Because the main energy of a flare is from its peak, the time
resolution is not enough and the profile is steep. Those characters
make the peak much smaller than the true value. We compare the
energy ratios between LC data and SC data, and find that LC data
have underestimated energies by 25\% with about 60\% errors.

(3) LC data also underestimate the flare amplitude by 60\% with
about 20\% errors, which has similar reasons as the estimation of
the flare energy.

(4) The durations of LC flares are overestimated. This conclusion is
determined by the steep profile of the flare and the time resolution
of LC data. Because a short, burst signal can enhance the average
value of the surrounding flux. Moreover, their deviations are
largely scattered, which depend on the fit of baselines. This is
because a flare has an exponential decay, its end point is close to
the quiescent flux, and decrease slowly. Slight changes of baselines
could affect the determination of the end point. LC data have
overestimated durations by about 50\%, and the errors are about
50\%.

It is undoubted that energies and amplitudes of LC data are
underestimated compared with the SC data, and durations are
overestimated. However, the limitation of LC data may only be the
reflection of the tendency of the deviations. The further analyses
demonstrate that it is hard to make corrections for each star by LC
data, because LC data cannot capture the detailed profiles of
flares. A complicate profile of a flare may results in a large
dispersion. For example, the SC data may have multiple peaks, or
multiple flares compared with the corresponding LC flare
\citep{Davenport2014,Balona2015,Pugh2016}. The dispersion of those
relations reflects the diversity of SC flares. After the comparisons
and the corrections, we suggest LC data can generally capture the
basic properties of flares.

 Many small flares in LC data are probably artifacts. Deviations of flare properties are determined by flare
profiles. Those results may affect some important conclusions made
by LC data, e.g., the proportion of flare stars
\citep{Balona2015,Davenport2016,Yang2017}, the FFD given by LC data
\citep{Shibayama2013,Wu2015}, the nature of superflares
\citep{Kitze2014,Wich2014}, the strength of the magnetic field of
the flaring stars \citep{Balona2015,Yang2017}, and the study of
stellar activity \citep{Yang2017}, all of which should be considered
in the future work.

\acknowledgements
\begin{acknowledgements}

We sincerely thank the anonymous referee for the very helpful
constructive comments and suggestions, which have significantly
improved this article. We acknowledge support from the Chinese
Academy of Sciences (grant XDB09000000), from the 973 Program (grant
2014CB845705), and from the National Science Foundation of China
(grants NSFC-11333004/ 11425313). E.Q. acknowledges support from the
National Science Foundation of China (grants NSFC-11773037). The
paper includes data collected by the Kepler mission. Funding for the
Kepler mission is provided by the NASA Science Mission Directorate.
All of the data presented in this paper were obtained from the
Mikulsk Archive for Space Telescopes (MAST). STScI is operated by
the Association of Universities for Research in Astronomy, Inc.,
under NASA contract NAS5-26555. Support for MAST for non-HST data is
provided by the NASA Office of Space Science via grant NNX09AF08G
and by other grants and contracts
\end{acknowledgements}

\section*{Appendix}

\begin{figure}[!htb]
\center
\subfigure{\includegraphics[width=0.45\textwidth]{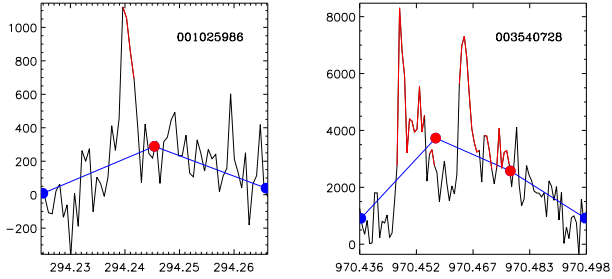}}
\hspace{2 mm}
\caption{\textbf{Light curves of two flares on KIC 1025986 and KIC
3540728}. The black lines are the detrended flux in SC data, and the
red lines are the flares in SC data. The blue lines are the
detrended flux in LC data, and the red points are the flares in LC
data. \textbf{It should be noted that one LC flare may correspond to
multiple SC flares as shown in the right panel, which is due to the
low resolution of LC data.}} \label{figapp}
\end{figure}

 For comparison, each flare includes LC and SC data is
plotted together. All images of flares are available only online. An
example is shown in Figure~\ref{figapp}. We suggest interested
parties view those images to have a direct impression of this work.

\end{document}